\documentstyle[multicol,aps,prb]{revtex}
\begin{document}

\draft

\title{
Flat-band ferromagnetism induced by off-site repulsions
}

\author{
Ryotaro Arita
}
\address{Department of Physics, University of Tokyo, Hongo,
Tokyo 113, Japan}
\author{
Yukihiro Shimoi
}
\address{Electrotechnical Laboratory, 1-1-4 Umezono, Tsukuba 305, 
Japan}
\author{
Kazuhiko Kuroki and Hideo Aoki
}
\address{Department of Physics, University of Tokyo, Hongo,
Tokyo 113, Japan}
\date{\today}

\maketitle

\begin{abstract}
Density matrix renormalization group method is used to 
analyze how the nearest-neighbor 
repulsion $V$ added to the Hubbard model on 
1D triangular lattice and a railway trestle 
($t$-$t'$) model will affect the electron-correlation 
dominated ferromagnetism arising from the interference (frustration).  
Obtained phase diagram shows that 
there is a region in smaller-$t'$ side
where the critical on-site repulsion above which the system
becomes ferromagnetic is reduced when the off-site repulsion 
is introduced.
\end{abstract}

%\medskip

\pacs{PACS numbers: 71.10.-w,71.10.Fd,75.10.Lp}

\begin{multicols}{2}
\narrowtext

The itinerant-electron ferromagnetism arising from electron 
correlations has a long history, 
dating back to the works by 
Hubbard\cite{Hubbard}, Kanamori\cite{Kanamori}
and Gutzwiller\cite{Gutzwiller} for the Hubbard model.  
It has become increasingly clear that the criterion for the 
ferromagnetism is a formidable question when the correlation effect 
is fully taken into account.  
There is a rigorous work by Nagaoka\cite{Naga} in the limit of infinite 
repulsion and infinitesimal doping, but the ferromagnetism 
there is singular in that the spin stiffness vanishes as an 
inverse system size\cite{KAnaga}.

Recently, a new light has been shed on the problem, 
when Lieb\cite{lieb}, and later Mielke and Tasaki\cite{MieTasa}, 
have shown that 
flat (dispersionless) bands in the one-electron band structure are good news 
for the ferromagnetism.  
Remarkably, we can show that the correlation rigorously 
guarantees the ferromagnetism for arbitrary strength of the repulsion $U$ 
when the flat band is half-filled.  

Occurrence of flat bands requires two classes of special 
lattice structures.  Lieb's model exploits the lattices 
that have different numbers of sublattice sites, while 
Mielke's and Tasaki's model provide the flat bands from interferences 
between the nearest-neighbor transfer $t$ and 
more distant transfers $t'$, which are assumed to have 
sizeable magnitude ($t' \simeq t$).  
Kusakabe and Aoki\cite{Kusakabe} have shown that the ferromagnetism 
is indeed stable, since, first, the spin stiffness is finite 
both in $U/t \rightarrow 0$ and $\rightarrow \infty$ limits 
despite an apparent lack of relevant energy scales, and 
second, the 
magnetism survives for finite dispersions.  
Tasaki\cite{Tasaki} has shown exactly that the magnetism is not 
destroyed for some class of 
dispersive bands for sufficiently large $U$.

These systems should be regarded as insulators 
because half-filled flat bands are considered there. 
Whether a ferromagnetic ground state persists in {\it metallic phases}, 
i.e., for non-half-filled bands, is of great interest, since 
this will imply an itinerant ferromagnetism. 
Penc {\it et al}\cite{Penc} studied one dimensional(1D) models such as
triangles connected linearly,
which may be thought of as a strip cut out from Kagom\'{e} 
lattice, a realization of Mielke's model\cite{kagome}.
Intuitively, a single triangle alone has a frustration in 
spin configurations, which results in an effectively ferromagnetic 
exchange interaction.  
Penc {\it et al} have shown, 
using an effective Hamiltonian in some limit 
($U/t \rightarrow \infty$, etc), 
that the 1D triangular lattice (Fig.1b) also has a ferromagnetic effective
coupling between spins.
The system is then expected to show ferromagnetism
for general electron density in this limit.  
Sakamoto and Kubo\cite{Sakamoto} studied 
with the density matrix renormalization group (DMRG)\cite{WhiteDMRG} 
whether 
the ferromagnetic ground state survives finite dispersions when
hole is doped to the nearly flat band.

On the other hand, Daul and Noack\cite{Daul} numerically determined the 
phase diagram of a 1D Hubbard model having both
nearest- and next-nearest-neighbor transfers 
($t$-$t'$ Hubbard model), which is topologically equivalent to a 
railway trestle (Fig.1c).  
They have found a large ferromagnetic region 
for finite densities and finite on-site interactions.
While a 1D triangular lattice is a two-band system with the 
lower band being flat, the trestle 
has a single band, but its bottom can be 
nearly flat depending on the value of $t'$. 
Hence we may include this ferromagnetism in 
the flat-band ferromagnetism in a broader context.
\cite{MHcomm}

In the studies mentioned above, only the on-site repulsion($U$) is considered.
In real materials, however, 
there are additional terms of appreciable strengths such as 
nearest-neighbor 
charge-charge interaction($V$), bond-charge 
interaction($X$), exchange interaction($F$), or onsite
pair-hopping($F'$), which have been investigated\cite{Vollhardt2}.
For itinerant ferromagnetism, Kollar, Strack and Vollhardt have 
derived sufficient conditions for realizing a ferromagnetic ground state
for general lattices with one hole in a half-filled band (generalized
Nagaoka's case). They conclude that $F$ is important in stabilizing
ferromagnetism for finite $U$, 
while for the special case of $X=t$ the ferromagnetism is stable for $F=0$.

The purpose of the  present paper is to 
study the effect of the nearest-neighbor repulsion
$V$ for the flat-band ferromagnetism as exemplified by the 
1D triangular lattice or trestle.  
The reason why we focus on the effect of $V$ is that
$V$ usually takes the largest value among $V, X, F, F'$, 
playing an important role in such materials as organic compounds.
Physically, the repulsion $V$, despite being 
a charge-charge interaction, can affect
the magnetism through Pauli's exclusion that makes the 
interaction effectively spin-dependent, but 
how the effect is exerted is a non-trivial question.  
Here we have employed DMRG, a powerful method for investigating strongly 
correlated 1D systems, to obtain 
the phase diagram against $U$ and $t'$.
We have found that 
the ferromagnetic region shifts to the smaller-$t'$ side 
when $V$ is switched on.

It is heuristic to start with a single triangle 
consisting of sites 1,2,3 (Fig.\ref{pic}a).
The Hamiltonian is given by
\begin{eqnarray*}
{\cal H}&=&-t\sum_{i=1}^{2}\sum_{\sigma}
(c_{i \sigma}^{\dagger}c_{i+1,\sigma}+{\rm H.c.})\\
&&+t'\sum_{\sigma}
(c_{2j-1,\sigma}^{\dagger}c_{2j+1,\sigma}+{\rm H.c.})\\
&&+U\sum_{k=1}^{3}n_{k \uparrow}n_{k \downarrow}
+V\sum_{i=1}^{2}n_{i}n_{i+1}
\end{eqnarray*}
in standard notations, 
where $t$ is the transfer between 1,2 and 2,3, 
$t'$ the transfer between 1,3 with $j=1$, and 
$U$ the on site repulsion.  
The nearest-neighbor repulsion $V$ is assumed to act 
between 1,2 and 2,3 only.  
Hereafter we take $t=1$ as a unit of energy and 
consider the case of $t'>0$, which favors the occurrence of flat bands.

In Fig.\ref{toy1}, we show for two electrons on a triangle 
the difference in energy, 
$\Delta\equiv E_{\rm t}-E_{\rm s}$, 
between the lowest spin-triplet state 
($E_{\rm t}$) and the lowest spin-singlet state($E_{\rm s}$) 
as a function
of $V$ for $U=10$ with $t'$ varied from 0.15 to 0.25.  
Unexpectedly, the curves can be non-monotonic: 
For $t'=0.2$ the ground state, which is a spin-singlet at $V=0$, 
becomes a triplet around $V=5$, then re-enters into a singlet for $V>7$.

The curious behavior can be understood intuitively as follows. 
Let us assume that the effect of the repulsion $V$ can be taken into 
account by reducing $t$ to a smaller $t_{\rm eff}$, because
$V$ reduces the amplitude of an electron residing on the top site 1.
This is reminiscent of the approximation often adopted in the 
$t$-$J$ model, where the effect of the infinite on-site repulsion is
taken into account by a reduction in $t$.
When we vary $t'$ in the absence of $V$,
the high-spin state becomes 
most stable ($\Delta$ takes its minimum) at $t'=t$ as 
shown in Fig.\ref{toy2} for $U=10$.  
In the above argument $V=5$ should then correspond 
to the case where $t$ is reduced to $t_{\rm eff} \simeq t' (=0.2$ here).

Now, an obvious question is: can this ferromagnetism induced by 
a repulsion survive when we connect the triangles into a 1D chain?  
So we have applied the DMRG method
to the Hubbard model on a 
1D triangular chain and to the trestle 
to look into such a possibility. 

The phase boundary between the ferromagnetic and 
paramagnetic phases can be determined 
in finite size systems as follows.\cite{Daul}
For $V\neq 0$, 
we first calculate the ground-state energy ($E_{\rm FP}$) for
fully spin-polarized state with DMRG.  
There the Hilbert space is much smaller than 
that of spinful electrons and the energy can be obtained 
very accurately.
For $V=0$, the ground state energy can be readily obtained, 
since spinless fermions do not feel $U$.
Next we calculate with DMRG the ground-state energy ($E_{\rm G}$) 
of a system having equal numbers of up-spins and down-spins 
(with $S_z=0$) to compare with $E_{\rm FP}$.  
Alternatively we can directly calculate the total spin from
\[
{\bf S}^2_{\rm tot}=\sum_{i,j}\langle {\bf S}_i\cdot  {\bf S}_j\rangle.
\]From these independent 
methods we can determine whether the ground state
is ferromagnetic.  
In our calculation the two methods gave the same result.

The calculation has been performed mainly for 
the number of atoms $L=39$ 
with an open boundary condition, 
and the convergence with respect to the system size has been confirmed 
by extending $L$ to 59.  
We have kept up to 120 states
per block at each step.\cite{comment} 
Using the finite-size algorithm, we swept 
the system about ten times.
We stored the density matrix at each step to construct good initial
vector for each superblock diagonalization\cite{acDMRG}.
We have obtained accurate wave functions 
with inverse iteration and conjugate-gradient optimization 
after each diagonalization.  

Let us first look at the 1D triangular chain, 
for which the one-electron energy band has a perfectly flat branch when 
$t'=t/\sqrt{2}$.
The Hamiltonian is given as before, 
where we have now $1\leq i \leq L-1$,
$1\leq j \leq (L-1)/2$, $1\leq k \leq L$.  
We have found that the value of $U$ required for 
the ferromagnetic ground state can indeed be 
reduced for nonzero $V$
for appropriate values of parameters, 
i.e., the band filling $n$=(number of electrons)/(number of 
sites)$<0.4$ and $t'<0.5$.  
A typical result is displayed in Fig.\ref{tri}, where 
we plot $\Delta \equiv E_{\rm FP}-E_{\rm G}$ as a function of $U$ 
for $n=0.2$ and $t'=0.31$.  
The one-electron band structures are depicted in the inset of the figure. 
We can see that the ground state becomes ferromagnetic
for $U>5$ for $V=0$, while the onset of 
the ferromagnetism is reduced to $U=4$ for $V=2$.  
Thus there is a parameter region in which the 
ferromagnetism is realized 
only when the off-site repulsion exists, 
as in the case of a single triangle.

The result that this phenomenon occurs only for 
smaller band fillings ($n<0.5$) may be 
understood as follows.  
In the above argument for a single triangle, we have 
assumed two electrons.
If we envisage that we require at least one hole per triangle 
(i.e., one hole per unit cell that consists of two sites) for 
the triangle lattice 
in a similar manner, this precisely amounts to $n<0.5$.

Let us next consider the $t$-$t'$ Hubbard model(fig. \ref{pic}c).
The Hamiltonian is given as
\begin{eqnarray*}
{\cal H}&=&-t\sum_{i,\sigma}
(c_{i \sigma}^{\dagger}c_{i+1 \sigma}+{\rm H.c.})
+t'\sum_{i,\sigma}
(c_{i \sigma}^{\dagger}c_{i+2 \sigma}+{\rm H.c.})\\
&&+U\sum_{i}n_{i \uparrow}n_{i \downarrow}
+V\sum_{i}n_{i}n_{i+1}.
\end{eqnarray*}
In Fig.\ref{zigzag} we again plot 
$\Delta\equiv E_{\rm FP}-E_{\rm G}$ as a function of $U$ 
for $t'=0.16$, $n=0.4$, and $V=0$ or $4$.  
It can be seen that the critical value, $U_{\rm c}$, above 
which the ferromagnetism appears is reduced from $U_{\rm c}=7$ 
down to $U_{\rm c}=5$ for $V=4$.  
When we make the off-site repulsion too large ($V=20$), 
the ferromagnetic ground state is washed away 
at least for $U<8$.
Thus, as in a triangle and in a triangle chain, 
an intermediate value of $V$ makes the
ground state to be ferromagnetic 
for appropriate values of $t', n$ and $U$.  

Figure \ref{phase} is the full phase diagram on the $U-
t'$ plane.  From this we can see that the ferromagnetic 
region, which has a concave boundary for $V=0$, 
{\it shifts} to the lower side of $t'$ when $V$ is introduced.  
The reduction of $U_{\rm c}$ found above is one manifestation of 
this.  
Intuitively, the reason why we have a concave ferromagnetic 
boundary with a minimum around $t' \simeq 0.36t$ for $V=0$ 
is that the ferromagnetism becomes most favorable 
when the 
one-electron band, displayed in the accompanied panels 
in the figure, approaches a flat band.  
Namely, the band in the $t-t'$ model cannot become exactly 
flat, but can become nearly so for $t' \simeq 0.36t$, which is in 
between the single-minimum band and the double-minimum band.  
The shift of this concave curve with $V$ may be understood qualitatively from 
the above argument that $V$ effectively reduces $0.36t$ into 
$0.36t_{\rm eff}$. 

For the generalized Nagaoka's ferromagnetism, 
a sufficient condition for a reduction of $U_c$ has been 
shown by Kollar {\it et al}\cite{Vollhardt2}, 
where for $F \neq 0$ or $X=t$ the condition on $U$ is
relaxed when we take an appropriate value of $V$.
The present result implies 
that for small $t'$,
off-site repulsion can stabilize the flat-band ferromagnetism as well.

To summarize, we have found 
for (nearly) flat bands arising from frustration 
(1D $t$-$t'$ model and triangular lattice) 
a region in the phase diagram where the critical on-site
repulsion for ferromagnetism is reduced 
when we introduce an off-site repulsion.
Recently, ferromagnetism in three-dimensional 
lattice structures (bcc, fcc, etc) 
has received attention\cite{Nolting,Hanisch,Ulmke}.  
Since network formed by the transfers in 
bcc and fcc lattices are also frustrated in that 
they comprise triangles,
it may be interesting to analyze the effect of $V$ 
in these systems.

R.A. is grateful to K. Kusakabe and H. Sakamoto for 
useful discussions. Computations were done on
SR2201 at the Computer Center of the University of Tokyo
and FACOM VPP 500/40 at the Supercomputer Center, Institute
of Solid State Physics, University of Tokyo.
%%%%%%%%%%%%%%%%%%%%  References %%%%%%%%%%%%%%%%%%%%%%%%%%%%%%%%

%%%%%%%%%%%%%%%%%%%% Figure Captions %%%%%%%%%%%%%%%%%%%%%%%%%%%%%%%
\begin{figure}
\caption{
A single triangle (a), 1D triangular lattice(b), and 
the railway trestle (or $t$-$t'$) model(c).
}
\label{pic}
\end{figure}

\begin{figure}
\caption{
The difference in energy, $E_{\rm t}-E_{\rm s}$, between the lowest triplet
state and the lowest singlet state for a single triangle 
plotted as a function
$V$ with $U=10$ for $t'$ varied from 0.15 to 0.25.
}
\label{toy1}
\end{figure}

\begin{figure}
\caption{
The difference in energy, $E_{\rm t}-E_{\rm s}$, between the lowest triplet
state and the lowest singlet state for a single triangle 
plotted as a function
$t'$ for $U=10$ with $V=0$.
}
\label{toy2}
\end{figure}

\begin{figure}
\caption{
The difference in energy ($\Delta$) between the lowest 
ferromagnetic state and the ground-state for a 
1D triangular lattice
plotted as a function of $U$ for
$t'=0.31$, $n=0.2$ with $V=0$ or $2$.  
Solid circles represent the result for 39 sites, diamonds for 59 sites, 
while the arrows indicate $U_{\rm c}$.
Inset shows one-electron band structures, where 
the horizontal dashed line indicates the Fermi level.
}
\label{tri}
\end{figure}

\begin{figure}
\caption{
A similar plot as in the previous figure 
for the $t$-$t'$ Hubbard model 
for $t'=0.16$, $n=0.4$ with $V=0$ or $4$.
}
\label{zigzag}
\end{figure}

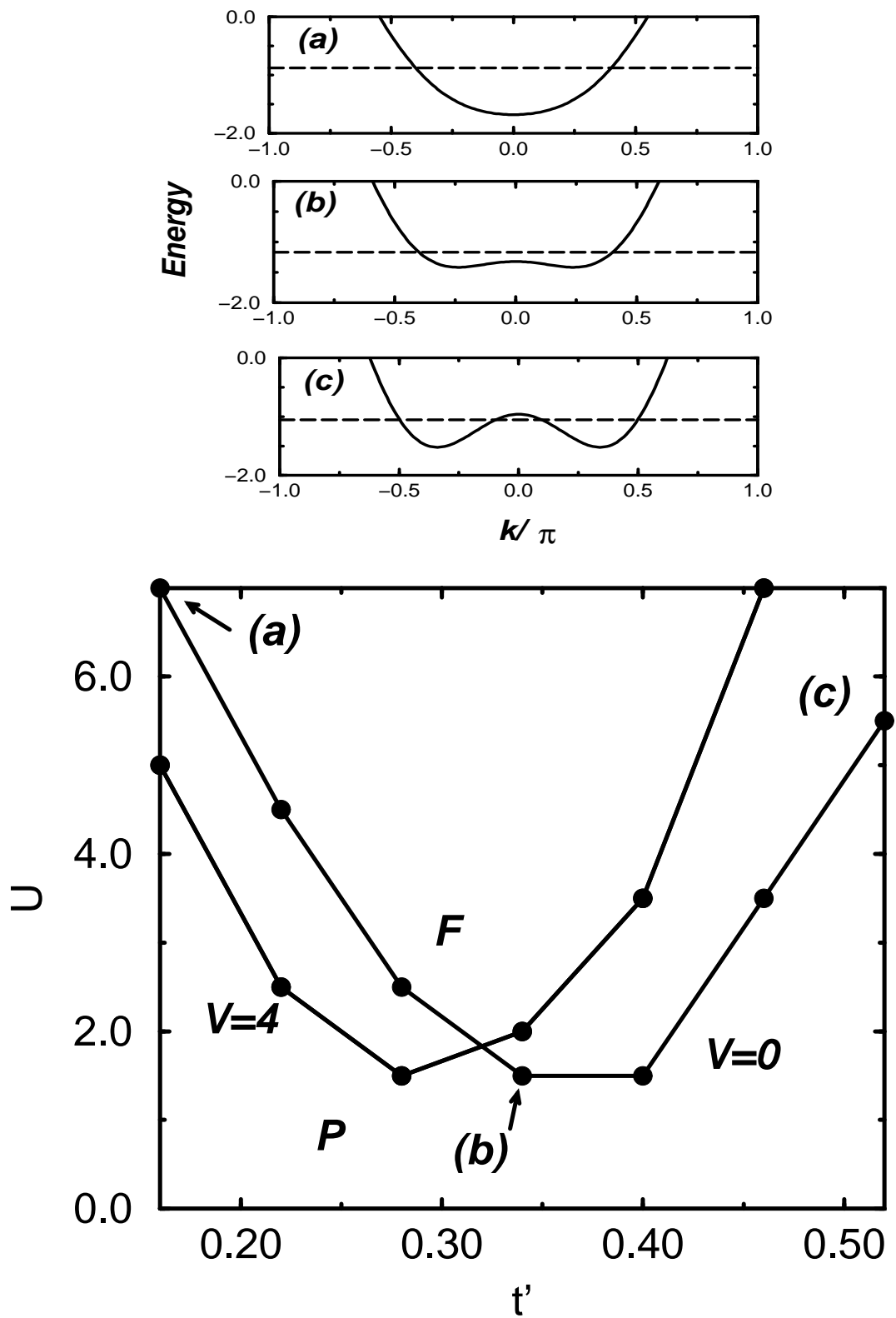
\begin{figure}
\caption{
The phase diagram (P: paramagnetic, F: ferromagnetic) against $U$ and $t'$ 
for the $t$-$t'$ Hubbard model 
for $n=0.4$ with $V=0$ or $4$.  
One-electron band structures are displayed in the accompanied panels 
for three typical values of $t'$ labeled with (a),(b),(c), where 
the horizontal dashed lines indicate the Fermi level.
}
\label{phase}
\end{figure}

\end{multicols}
\end{document}